\documentclass[fleqn,twoside,onecolumn,nofootinbib,showkeys,12pt]{revtex4}

\usepackage{graphicx}% Include figure files

\usepackage{bm}% bold math

  \newfont{\cyrfnt}{wncyr10 scaled 1120}

\begin{document}

\begin{center}

{\large \bf Thermodynamically Anomalous Regions As A Mixed Phase Signal}

\vspace*{11mm}

{\bf K.A. Bugaev$^{1,*}$, A.I. Ivanytskyi$^1$, D.R. Oliinychenko$^{1,2}$,  V.V.~Sagun$^1$,  
I.N.~Mishustin$^{2, 3}$, D.H. Rischke$^4$, L.M. Satarov$^{2, 3}$ and  G.M.~Zinovjev$^1$}

\vspace*{5.5mm}

{\small

$^1${Bogolyubov Institute for Theoretical Physics, of the National Academy of Sciences of Ukraine, Metrologichna str. 14$^B$, Kiev 03680, Ukraine}\\

$^2${FIAS, Goethe-University,  Ruth-Moufang Str. 1, 60438 Frankfurt upon Main, Germany}\\

$^3${Kurchatov Institute, Russian Research Center, Akademika Kurchatova Sqr., Moscow, 123182, Russia}\\

$^4${Institute for Theoretical Physics, Goethe-University,  Ruth-Moufang Str. 1, 60438 Frankfurt upon Main, Germany}\\
}

$^*${e-mail: bugaev@th.physik.uni-frankfurt.de}

%%{Quark deconfinement, quark-gluon plasma production, and phase transitions}
%%Pacs{25.75.-q}{Relativistic heavy-ion collisions}
%%Pacs{25.75.Dw}{Particle and resonance production}

\end{center}

{\bf Abstract.}
Using the most advanced  model of the hadron resonance gas  we reveal, 
at chemical freeze-out,  remarkable irregularities such 
as an abrupt change of the effective number of degrees of freedom 
and plateaus in the collision-energy dependence of the entropy per baryon, total pion 
number per baryon, and thermal pion number per baryon at 
laboratory energies 6.9--11.6 AGeV. 
On the basis of the generalized shock adiabat model we show  that these plateaus give 
evidence for the thermodynamic anomalous properties of the mixed phase 
at its boundary to the quark-gluon plasma (QGP).  A new signal for QGP formation 
is suggested and justified.
\\


%% Beginning         08.08.2013              Kyrill       Horn_Puzzle_1

\section{Introduction} In the last thirty years of searching for the QGP in heavy-ion collision
experiments many signals of its formation were 
suggested, but neither direct evidence for the QGP nor a clear signal of a 
QGP-hadron mixed phase have been observed 
so far. Although some irregularities, known in the literature as the Kink \cite{Kink}, the 
Strangeness Horn \cite{Horn} and the Step \cite{Step}, were observed and are considered  to 
be signals of the onset of deconfinement \cite{Gazd_rev:10}, their relation to the 
QGP-hadron mixed phase is far from being clear.
Therefore,  additional  and independent justification of these irregularities is required. 
This task is rather important in  view of the planned heavy-ion collision  experiments at 
JINR-NICA  and  GSI-FAIR.
Evidently, searching for other irregularities and signals of mixed-phase formation and 
their justification is no less significant, but until recently such efforts were not  very  
successful,  since they require a realistic model  which is able to
accurately  describe the existing experimental data and, thus, provide us with 
reliable information about the late  stages of the 
heavy-ion collision process. 

\begin{figure}[t]
\centerline{\includegraphics[width=91.8mm]{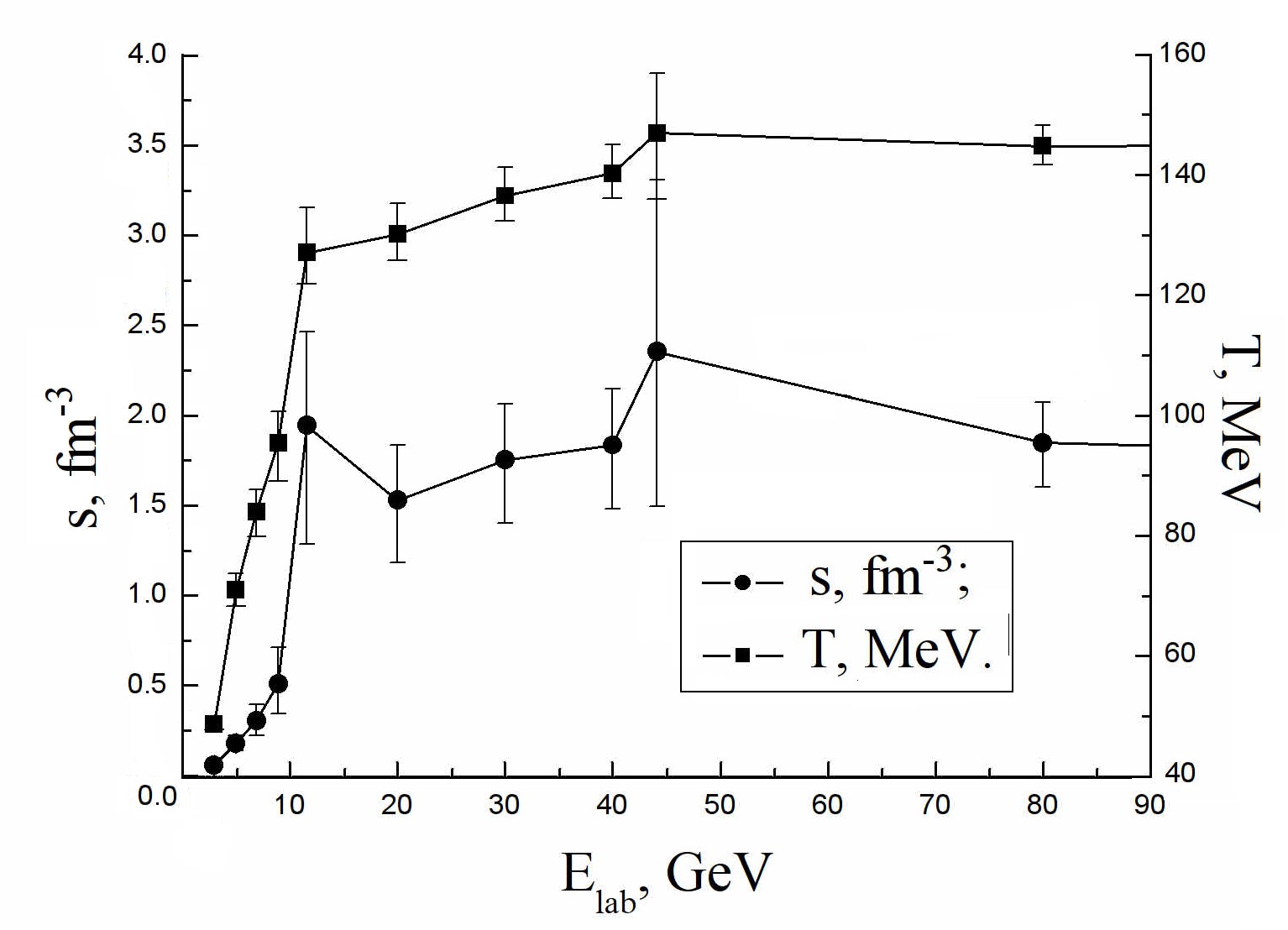}
}
 \caption{Energy dependence of entropy density (circles) and temperature (squares) at 
 chemical freeze-out extracted in \cite{KAB:13} from hadron multiplicities measured in 
 heavy-ion collisions.
}
  \label{fig1}
\end{figure}

The recent extensions \cite{Oliinychenko:12,KAB:13,KAB:gs,NICAWP} of the hadron 
resonance gas model 
\cite{KABCleymans:93,KABCleymansFO,KABpbm:02,KABAndronic:05,KABAndronic:09}
provide us with the most successful description  of available hadronic multiplicities measured
in heavy-ion collisions  at AGS, SPS, and RHIC energies. 
The global values of $\chi^2/dof \simeq 1.16$  
and  $\chi^2/dof \simeq 1.06$  achieved, respectively, in \cite{KAB:13} and  \cite{KAB:gs}  for 
111 independent multiplicity ratios measured at fourteen values of collision energy  
give us confidence that  the irregularities shown in Figs.\ 1 and 2  are not 
artifacts of the model and indeed reflect  reality. From Fig.\ 1 one can see
that the entropy density $s$ increases by a factor of 4
in a range of laboratory energies per nucleon $E_{lab} \simeq 8.9 - 11.6$  GeV. At the same
time, the chemical freeze-out (FO) temperature changes from 95 to 127 MeV and the
baryonic chemical potential $\mu_B$ drops from 586 to 531 MeV \cite{KAB:13}. 
In other words, for a 30\% increase in the laboratory energy  the ratio $s/T^3$ increases
by 70\%. A similar change  can be seen in 
the effective number of degrees of freedom in the same  energy range, cf.\ 
Fig.\ 2 for the chemical FO  pressure in units of $T^4$. Note that  a similar 
(and  a somewhat  stronger)  rapid change in the number of effective degrees of freedom 
is observed in   the most recent (former) version of  the hadron resonance gas model
\cite{KAB:gs,NICAWP} (\hspace*{-1.1mm}\cite{Oliinychenko:12,KABAndronic:05}).

\begin{figure}[t]
%
%\centerline{\includegraphics[height=67.1mm]{Fig2.jpg} 
\centerline{\includegraphics[width=110.0mm]{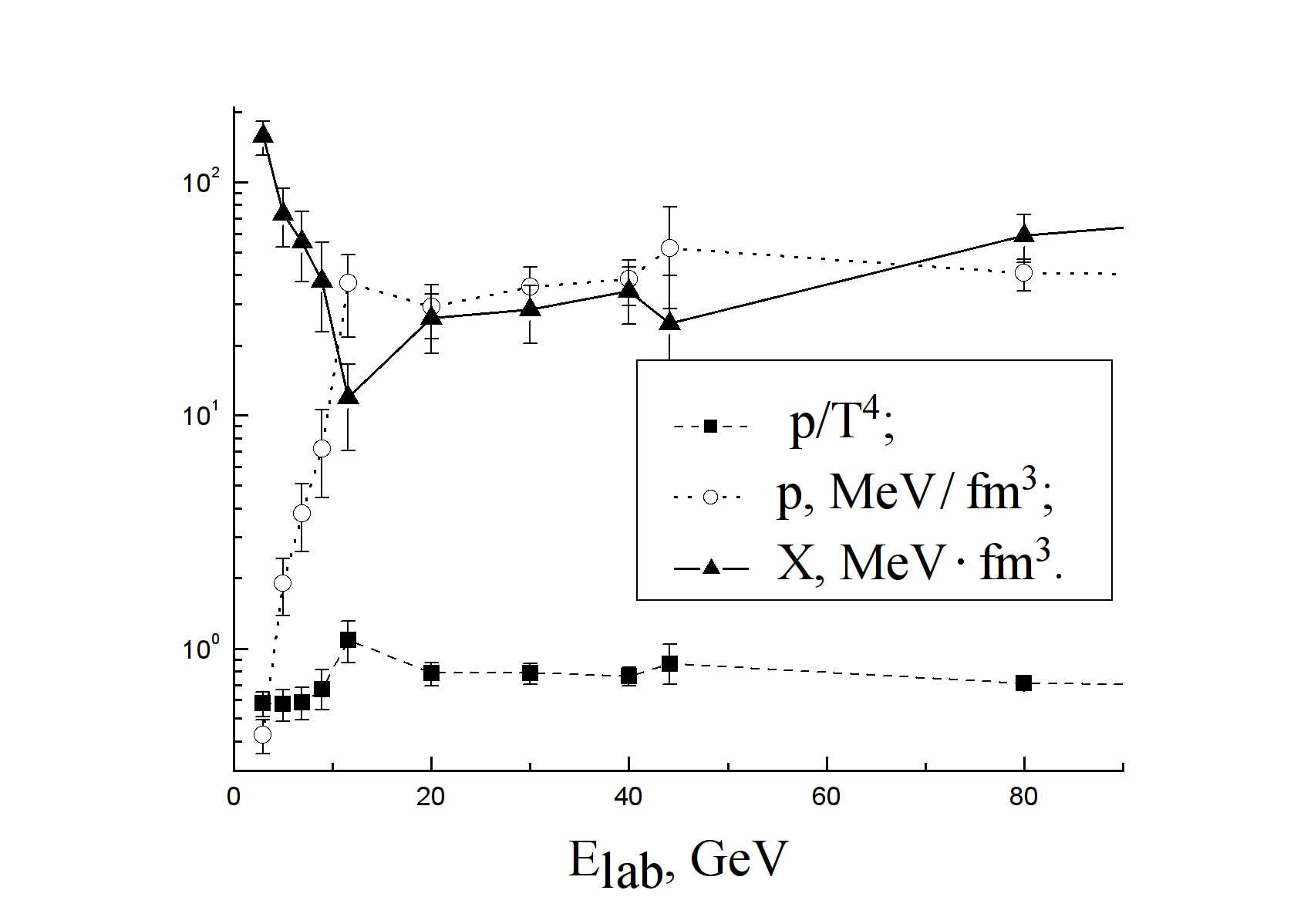}
}
 \caption{Energy dependence of the chemical freeze-out pressure (circles),  
 the effective number of degrees of freedom (squares),  
 and the generalized specific volume $X$  found by the model of   \cite{KAB:13}.
}
  \label{fig2}
\end{figure}

As one can see from Fig.\ 2, at chemical FO  a more dramatic change is  
experienced by  the so-called generalized specific volume 
$X=\frac{\varepsilon +p}{\rho_B^2}$, where
$\varepsilon$ is the energy density, $p$ the pressure,
and $\rho_B$ the baryonic charge density.  
It is remarkable that in all known examples of  equations of state (EOS)
describing the QGP-hadron transition a  local minimum 
in the energy dependence of the  $X$ values of matter described by the shock or generalized 
shock model
is observed right at the transition to the QGP, independently of whether this is a first-order 
phase transition \cite{KAB:89, KAB:89a,KAB:90, KAB:91} or a  
strong cross-over \cite{KAB:89, KAB:89a}.
Therefore, here we would like to reanalyze  the  generalized shock adiabat model developed
in \cite{KAB:89, KAB:89a, KAB:90, KAB:91, KAB:89b} in order to interpret the above 
irregularities and to verify the other signals of mixed-phase formation.
%%% suggested in \cite{KAB:89a,KAB:90,KAB:91}.

\section{Generalized Shock Adiabat Model}Such a model was developed in 
\cite{KAB:89, KAB:89a, KAB:90, KAB:91, KAB:89b} to extend the  compression shock model 
\cite{Mish:78, Horst:80, Kamp:83,Horst_PR:86, Barz:85} for regions of   matter with anomalous
thermodynamic properties. Similarly to nonrelativistic hydrodynamics \cite{Landavshitz}, 
in the relativistic case   the matter is thermodynamically normal, if the quantity 
$\Sigma \equiv \left(\frac{\partial^2 p}{\partial X^2} \right)^{-1}_{s/\rho_B}$ 
is positive along the Poisson adiabat. Otherwise, for $\Sigma < 0$, the 
matter has thermodynamically anomalous properties. The sign of $\Sigma$ 
defines the type of  allowed simple and shock waves: for  $\Sigma > 0$  
rarefaction simple waves and compression shocks  are stable.  In the case of an 
anomalous medium  compressional simple waves and rarefaction shocks  are stable. 
If both signs of $\Sigma$  are possible, then a more detailed investigation of the 
possible flow patterns is necessary \cite{KAB:89a,KAB:91}.  In fact, all known pure 
phases have thermodynamically normal  properties, whereas  anomalous properties may
appear at a first-order phase transition \cite{Zeldovich,Rozhd}, at its second-order
critical endpoint \cite{Landavshitz}, or for a fast cross-over \cite{KAB:89a}.

\begin{figure}[t]
\centerline{\includegraphics[width=84.0mm]{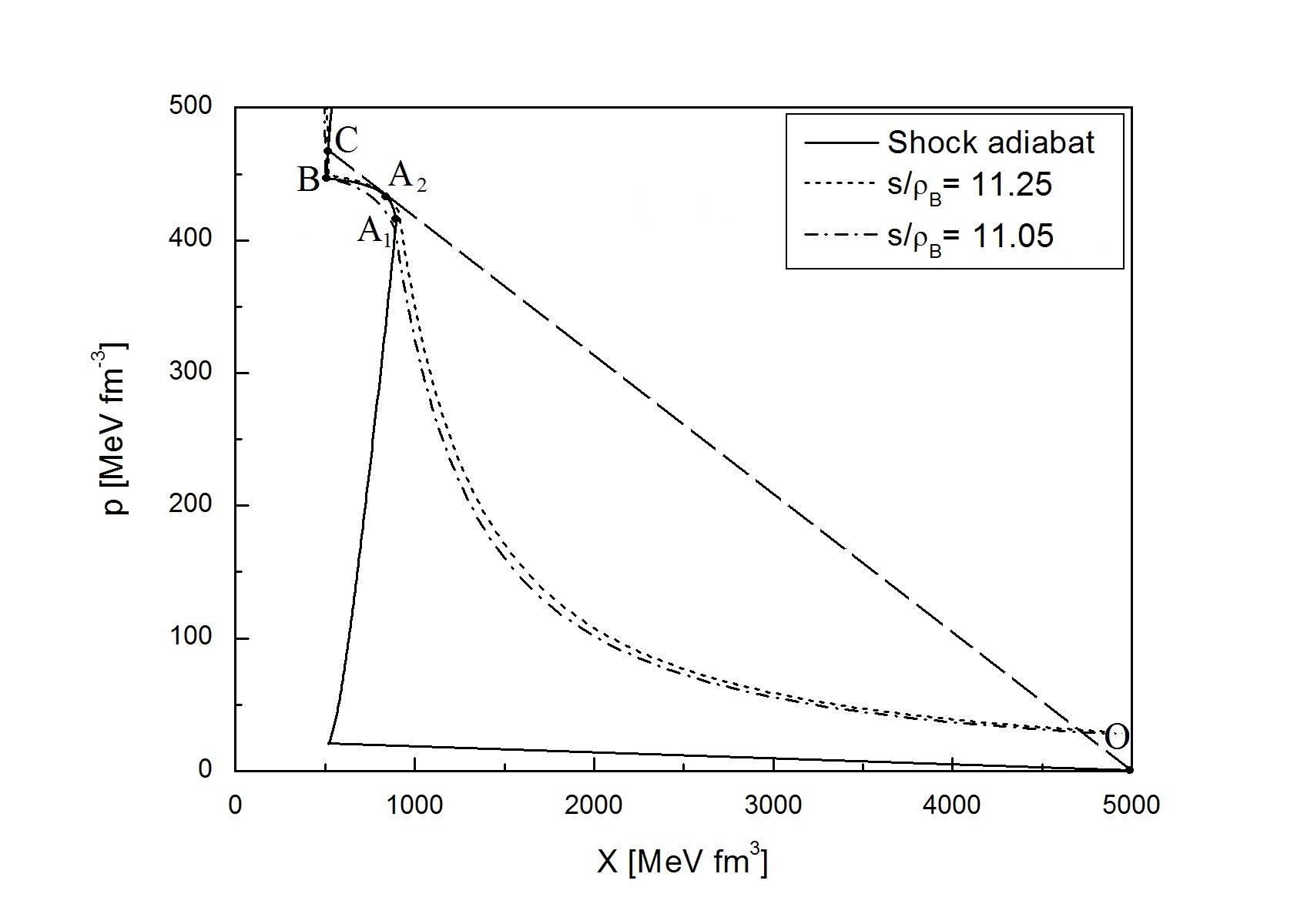}}
\vspace*{-3.3mm}
 \caption{The compression RHT adiabat OA$_2$BC (solid curve)  of W-kind  in the $X-p$ 
plane.  It is  calculated for an  EOS with first-order phase transition discussed in the text.  
The segments OA$_1$, A$_1$B, and BC of the adiabat 
correspond to the hadronic, mixed, and QGP phases, respectively.   
Shock transitions into the region of states  A$_2$BC are mechanically unstable. The tangent 
point A$_2$ to the shock adiabat is the Chapman-Jouguet point \cite{Landavshitz}.
The dotted and dashed-dotted curves show the Poisson adiabats with 
values of entropy per baryon specified in the legend.}
  \label{fig3}
\end{figure}

The compression shock model of central nuclear collisions 
\cite{Mish:78, Horst:80, Kamp:83,Horst_PR:86, Barz:85} allows one to determine the 
initial conditions for the subsequent hydrodynamic evolution. Such a picture of the 
collision process,  which neglects the nuclear transparency,  can be reasonably  well  
justified at intermediate collision energies per nucleon $1$ GeV $ \le E_{lab} \le $ 15 GeV.  
At laboratory  energies per nucleon  up to  20 GeV this model can  
be used for  quantitative estimates, while at  higher energies  it provides a 
qualitative description only.  In the center-of-mass frame of the two colliding nuclei  the 
initial moment of the collision can be considered as a hydrodynamic Riemann problem 
of an initial discontinuity. For normal media this  kind of initial discontinuity  leads to 
an appearance of two compression shocks that move in  opposite directions 
toward the vacuum, leaving  high-density matter at rest behind the shock fronts.
The thermodynamic  parameters $X, p, \rho_B$  of this compressed matter  
are related by  the Rankine-Hugoniot-Taub (RHT) adiabat \cite{Landavshitz} 
with uncompressed matter in the state  $(X_0, p_0,\rho_{B0})$, 
\begin{eqnarray}\label{EqI}
\rho_{B}^2 X^2 -  \rho_{B0}^2 X^2_0 = (p - p_0)\left( X + X_0  \right) \,. 
\end{eqnarray}
This equation follows from the usual hydrodynamic conservation laws of energy, momentum, 
and baryonic charge across the  shock front. 
The variable $X$   is convenient,  since with its help the conserved  baryonic current 
can be expressed as  
%
%%\begin{eqnarray}\label{EqII}
%
$j^2_B = - \frac{ p - p_0}{ X - X_0 }$,
%
%%\end{eqnarray}
%
i.e., in the $X-p$ plane the state  existing behind   the shock front is 
given by the intersection point 
of the  RHT adiabat (\ref{EqI}) and the straight line  with the slope $j^2_B$ 
known as the Raleigh line. To solve Eq.\ (\ref{EqI}) one needs to know  the EOS.  
Within the compression shock  model the laboratory energy per nucleon is
\begin{eqnarray}\label{EqII}
E_{lab} = 2 m_N \left[ \frac{(\varepsilon+p_0)(\varepsilon_0+p)}{(\varepsilon+p)(\varepsilon_0+p_0)} - 1 \right] \,,
\end{eqnarray}
where $m_N $ is the mean nucleon mass. A typical  example for the shock adiabat is shown in  
Fig.\ \ref{fig3}. As one can see from this  figure  the shock adiabat  in the pure 
hadronic and QGP phases  exhibits the typical (concave) behavior for a normal medium, while 
the mixed phase (the region A$_1$B)
in Fig.\ \ref{fig3} has  a convex shape which is typical for matter with 
anomalous properties. 
Until  now there is no complete understanding why in a phase-transition or cross-over  region 
matter exhibits  anomalous thermodynamic properties. In pure gaseous or liquid phases the 
interaction between the constituents  at short distances is repulsive and, hence, at high  
densities the adiabatic  compressibility of matter 
$- \left( \frac{\partial X}{\partial p}\right)_{s/\rho_B}$ usually decreases for 
increasing pressure, i.e.,  
$ \left(\frac{\partial^2 p}{\partial X^2} \right)^{-1}_{s/\rho_B} =  \Sigma >0$. In the mixed phase 
there appears another possibility to compress matter: by converting the less dense phase 
into the more dense  one.  As it was found  for several EOS with a first-order phase transition 
between hadronic gas and QGP, the phase transformation leads to an increase of the
compressibility in the mixed phase at higher pressures, i.e., to 
anomalous thermodynamic properties. 
The hadronic phase of the aforementioned EOS was described   by the Walecka model 
\cite{Walecka} and by a few  of its more realistic phenomenological generalizations 
\cite{KAB:91, Zimanyi, Satarov:11}.
The appearance of anomalous thermodynamic properties  for a fast cross-over can 
be understood similarly, if one formally considers the cross-over states  as a kind of 
mixed phase (but without sharp phase boundary), in which, however,  
none of  the  pure phases is able to completely dominate.

%%%\clearpage

\begin{figure}[t]
\centerline{\includegraphics[width=84.0mm,height=55mm]{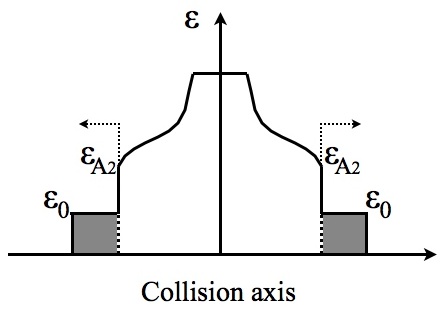}}
%
%
%%%\centerline{\includegraphics[width=84.0mm]{Fig_3b.jpg}}
%
 \caption{Sketch of a collision of  two nuclei (grey areas) where the
generalized shock adiabat states 
are above the Chapman-Jouguet point $A_2$. Two shocks 
 between  the states $\varepsilon_0 \rightarrow \varepsilon_{A_2}$ are 
 followed by compressional simple waves. The dashed arrows show the 
 direction of shock propagation.}
  \label{figN4}
\end{figure}

From Fig.\ \ref{fig3} one sees that  the presence of anomalous matter  
leads to mechanically unstable parts of  the RHT adiabat (segment $A_2BC$ in Fig.\ \ref{fig3}) 
which include states in the  mixed and the QGP phases. 
This is  a  model of W-kind \cite{Walecka, KAB:91, Satarov:11} and   its   RHT adiabat in the instability  
region should be replaced by the generalized shock adiabat \cite{KAB:89a,KAB:90,KAB:91}. 
In the region of instability  the shock wave for W-kind models has to be replaced 
by the following hydrodynamic  solution \cite{KAB:89a}: a shock between 
states O and A$_2$ (on the RHT adiabat shown in Fig.\ \ref{fig3}), followed by a compressional 
simple wave  (see Fig.\ \ref{figN4}); at higher energies this solution converts into  two 
compressional shocks and a compressional simple wave moving between them. 
A similar situation occurs in the case of a fast cross-over  
(see Figs.\ 3 and 4 in \cite{KAB:89a} for more details). 
An additional  solution of two compressional shocks following one after the other may appear, if 
all transitions to the mixed phase are unstable \cite{KAB:89a,Mish:78}. 

Shock transitions to mechanically unstable regions are accompanied by a
 thermodynamic instability, i.e., the entropy
 in such transitions decreases, while collision energy grows \cite{Zeldovich,Rozhd,KAB:89a}. 
 At the same time the mechanical stability condition of the  generalized shock adiabat 
 always leads to 
 thermodynamic stability of  its  flows. Or in other words,  along the correctly constructed  
 generalized shock adiabat  the entropy cannot decrease  \cite{KAB:89b}. 
Among the possible solutions mentioned above an important role is played by 
the combination of a
shock wave between the states O and A$_2$, followed by a simple wave starting in  
the  state A$_2$ and continuing to states located at the  boundary  between  the mixed 
phase  and the QGP \cite{KAB:89a}. 
For such a solution the entropy is conserved, i.e., the ratio of entropy density   
per baryon $s/\rho _B = const$, 
because the whole entropy production is generated by a shock OA$_2$ to
the Chapman-Jouguet point A$_2$ (see Fig.\ \ref{fig3}).
This  means that by increasing the collision energy one generates more 
compressed states which, however,  have 
the same value of $s/\rho _B$.

Based on this solution  a signal for mixed-phase formation was suggested, provided  
that  this instability of a W-kind 
model exists \cite{KAB:89a, KAB:90, KAB:91}. 
The important physical consequence of such an instability
is a plateau in the collison-energy dependence  of the total number of pions per baryon
produced in a 
nuclear collision, i.e., $\rho_\pi^{tot}/\rho_B (E_{lab}) \simeq const$  \cite{KAB:90, KAB:91}, 
provided by the entropy conservation during the subsequent expansion 
of the   hydrodynamic flow formed  by the generalized shock adiabat. 
Since the total pion multiplicity consists of thermal pions and the ones which appear 
from decays of hadronic resonances, in case of  the RHT adiabat instability  the  
number of thermal pions per baryon $\rho_\pi^{th}/\rho_B$  should also 
demonstrate a plateau or  a plateau-like  behavior with a small negative slope  
as  a function of collision energy. 

%%%\vspace*{2.2cm}
%%%KAB

\begin{figure}[t]
\centerline{\,\,\,\,\,\includegraphics[width=88.0mm,height=61.7mm]{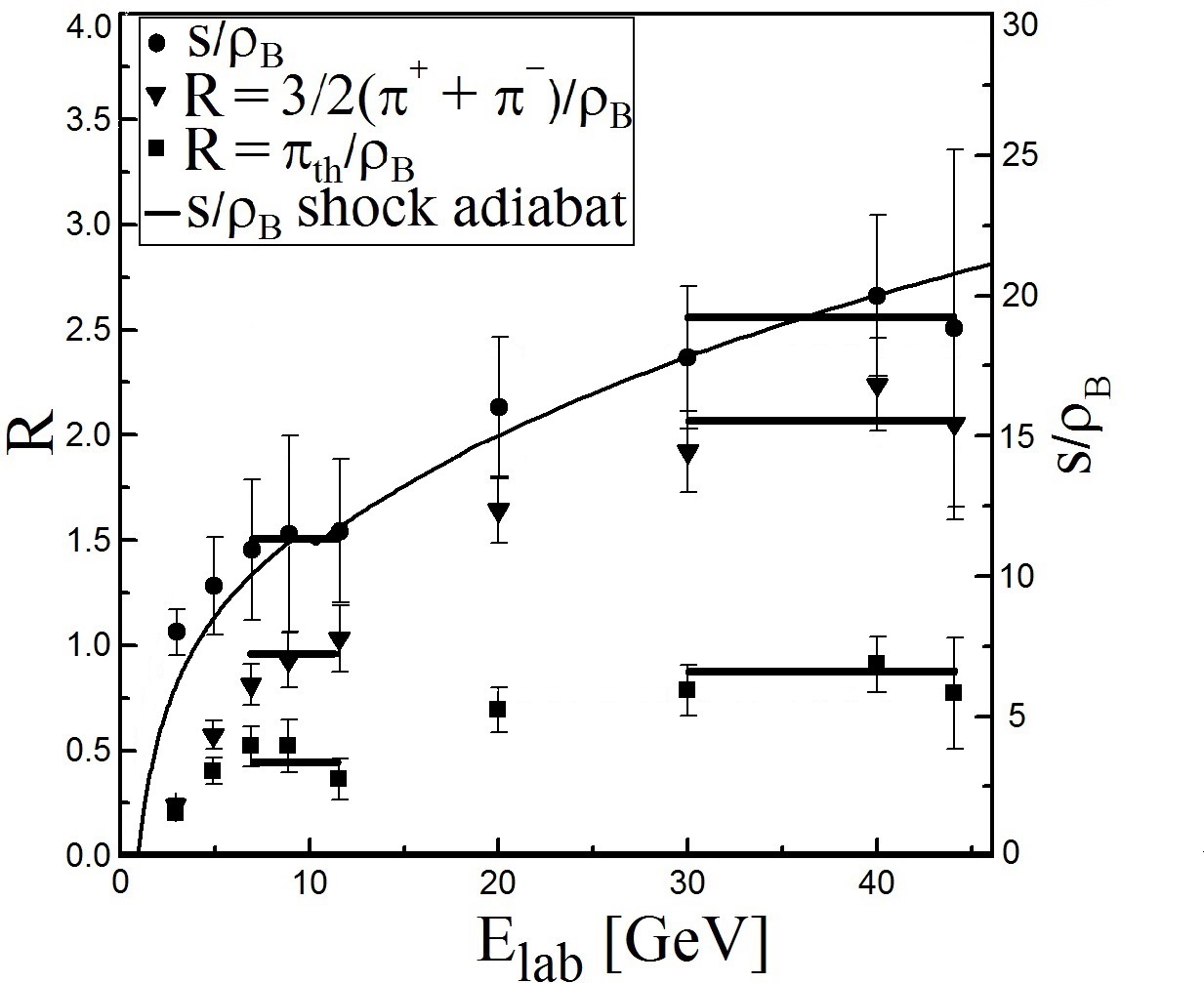}
}
 \caption{Energy dependence of the entropy per baryon  (circles), of  the thermal pion 
 multiplicity per baryon  (squares),  and  of  the  total pion multiplicity per baryon  (triangles) 
 found at the chemical freeze-out  within the realistic version of the hadron resonance gas 
 model  \cite{KAB:13}. The horizontal bars are found by minimizing $\chi^2/dof$ (see text). 
 The solid curve corresponds to the RHT adiabat shown in Fig.\  \ref{fig3}.
}
  \label{fig5}
\end{figure}

Note that  the proposal of possible appearances of plateaus  in the entropy per baryon 
and in the total pion number per baryon as functions of collision energy was  strongly 
criticized in the literature. In Fig.\ \ref{fig5} one can see all three 
plateaus  at the laboratory energies   $E_{lab} \simeq 6.9 - 11.6$ GeV, i.e.,  exactly where the 
other irregularities depicted in Figs.\ 1 and 2 occur.  All quantities  shown in Figs.\ 1, 2, and 5 
were found at chemical FO within the most realistic model of the hadron resonance gas  
 with multicomponent  hard-core repulsion 
\cite{KAB:13}, which not only  successfully describes 111 independent 
hadron multiplicity ratios measured for  
center-of-mass energies  $\sqrt{s_{NN}} = $ 2.7 -- 200 GeV, but also  correctly  
reproduces  the energy dependence  of the Strangeness Horn 
with a $\chi^2/dof \simeq 7.5/14$. 
 
It is, of course, possible that the anomalous properties of the mixed phase do not generate the 
mechanical instabilities for 
the shock transitions to this phase \cite{KAB:89a}, like it was found for the Z-kind  
hadronic EOS   \cite{Zimanyi}. Nevertheless, it was argued that even in the latter case 
a plateau-like structure in the entropy per baryon, and, hence, in  the thermal pion 
multiplicity per baryon, should also be  seen \cite{KAB:89a, KAB:90, KAB:91}.  

Now we are at a position 
to determine the parameters (individual heights for the same width)  of the plateaus in  the 
ratios   $s/\rho_B$,
$\rho_\pi^{th}/\rho_B$ and $\rho_\pi^{tot}/\rho_B$ shown in Fig.\ \ref{fig5}.
We investigated a few different schemes, but came to the  conclusion that a
3-parameter fit is the most reliable and simple one. Since we are searching for a plateau it is 
clear that its height $R_A$ should be the same for a given quantity 
$A \in \{s/\rho_B; \rho_\pi^{th}/\rho_B; \rho_\pi^{tot}/\rho_B\}$.  The width of all plateaus 
in the collision energy should also be the same, since they are generated by the 
same physical mechanism. Let  $i_0$ denote the beginning of plateau, 
while  $M$   denotes  its width.
Then one has to minimize 
\begin{equation}
\chi^2/dof= \frac{1}{3M-3}\sum_A \sum_{i=i_0}^{i_0+M-1}
\left(\frac{R_A-A_i}{\delta  A_i} \right)^2 
\label{Plat}
\end{equation}
for all  possible values of $i_0$ and $M > 1$.  Here the subscript  $i$ counts  the data 
points $A_i$ to be described, whereas  $\delta A_i$ denotes the error of the
 corresponding quantity $A_i$. 
We assume that the plateaus are correlated to each other, 
if $\chi^2/dof$ is essentially smaller than 1. 
Also from the practical point of view it is necessary to find the  set of  
maximally correlated plateaus for future experiments.
The height  of each plateau $R_A$ is found by minimizing $\chi^2/dof$ in  (\ref{Plat}) with 
respect to $R_A$ and one gets 
\begin{eqnarray}
\label{V}
R_A &= &\sum_{i=i_0}^{i_0+M-1}\frac{A_i}{\left(\delta A_i \right)^2}\biggl/
\sum_{i=i_0}^{i_0+M-1}\frac{1}{\left(\delta A_i \right)^2} \,.
\end{eqnarray}
As one can see from the table below minimal values of $\chi^2/dof < 0.2$ are 
reached for $M=2$, but these are not the widest plateaus. There exist two sets 
for  $M=3$ with  $\chi^2/dof \simeq 0.53$ for the low-energy plateaus (at $E_{lab} \simeq 6.9-11.6$ GeV)
and with  $\chi^2/dof \simeq 0.34$ for the high-energy plateaus (at $E_{lab} \simeq 30-40$ GeV). Precisely  these sets are 
depicted in Fig.\ \ref{fig5}, since we believe 
that the high-energy set of  $E_{lab} \simeq 20-40$ GeV  with the width $M=4$  should not be taken into account because  its 
value of  $\chi^2/dof \simeq 0.87$
is too close to 1 and, therefore,  such plateaus are not strongly correlated 
even for these huge error bars. 
\vspace*{-5mm}

\begin{table}[h]
\noindent\caption{Results of the 3-parameter fit.}\vskip-3mm\tabcolsep3.2pt
\begin{center}
\begin{tabular}{|c| c| c| c| c| c|}
\hline  
&                                                    \multicolumn{5}{c|}{Low energy minimum}  \\ \hline             
  $M$  & $i_0$ &  $R_{s/\rho_B}$ &  $R_{\rho_\pi^{th}/\rho_B}$ & $R_{\rho_\pi^{tot}/\rho_B}$ & $\chi^2/dof$  \\\hline 
     2     &    3    &      11.12988        &        0.52037                  &                 0.85683                  &     0.17811 \\\hline 
      3     &    3    &      11.31482        &       0.46128                   &                 0.89174                  &     0.53144  \\\hline  
 4     &    2    &      10.55597        &       0.43340                   &                 0.72523                  &     1.64913    \\\hline 
   5     &    2    &      11.53637        &       0.47009                   &                 0.84800                  &     4.45466   \\\hline    
 &                                           \multicolumn{5}{c|}{High energy minimum}                                                    \\ \hline
   
     2     &     8     &      19.80518       &         0.88229                 &                2.20373                    &      0.12751    \\\hline
  3     &   7     &      18.77659       &         0.83508                 &                2.05780                    &      0.34045           \\\hline 
  4     &   6     &      17.82325       &         0.77920                 &                1.87732                    &      0.87105   \\\hline 
    5  & 5    &     16.26105       &         0.64800                 &                1.62094                    &      3.72057      \\\hline          
\end{tabular}
\end{center}
\end{table}
\vspace*{-4.4mm}

%\section{New signal of QGP formation}
{\bf New signal of QGP formation.}--
Here we suggest a new signal indicating a boundary between the mixed phase and the QGP, 
which also  signifies  the  existence of  mechanical instabilities inside the mixed phase. 
This is an appearance of a local minimum of the generalized specific volume $X$ 
at chemical FO  as a function of the collision energy (see Fig.\ \ref{fig2}). 
Note that all stable RHT adiabats of Z-kind and all  unstable  RHT adiabats of W-kind and 
 the corresponding  generalized shock adiabats  with the QGP EOS 
of the  MIT-Bag model type \cite{MITBag} studied in 
\cite{KAB:89a, KAB:90, KAB:91, Satarov:11} demonstrate exactly the same   behavior. 
The physical origin for such a behavior is that for an increase in collision energy  
the entropy per baryon  
 and the temperature of the formed QGP (being a normal medium) increase as well, while the 
 baryonic density and the baryonic chemical potential are  steadily decreasing. 
 Hence in the QGP phase  the variable $X \equiv (T s/\rho_B + \mu_B)/\rho_B$  grows, if  the 
 collision energy increases. Intuitively, such a dependence  seems to be true for other QGP 
 EOS, if they correspond to a normal medium. On the other hand,  the behavior of the variable 
 $X$ inside the mixed phase with anomalous properties is opposite 
 and it does not depend on the stability or instability of  the shock transitions 
 to this region   \cite{KAB:89a, KAB:90, KAB:91}. On the basis of these arguments one  
 can understand the reason why  the boundary of the mixed phase and QGP corresponds  to a  
 local minimum of the $X$ variable along  the RHT (shown in Fig.\ \ref{fig3}) or generalized shock adiabat  and why it   is   also a minimum of $X$  as function of 
 collision energy \cite{KAB:89a, KAB:90, KAB:91, Satarov:11}.

In case of unstable shock transitions to the mixed phase,  the  unstable part of the mixed 
phase (segment A$_2$B in Fig.\ \ref{fig3}) should be replaced by the Poisson adiabat 
passing through the point A$_2$  (the dotted curve shown in Fig.\ \ref{fig3}).  
Consequently,  if the matter  formed  in a collision  expands isentropically after the 
shock OA$_2$ disappears, then it can be shown that  for 
the chemical FO   pattern depicted in Figs.\  \ref{fig1} and \ref{fig2} the minimum of the variable 
$X$  of the initial state
corresponds to a minimum of this variable at chemical FO, i.e., $\min\{X (E_{lab})\}$ 
corresponds to  $\min\{X^{FO} (E_{lab})\}$. Indeed, the final  states  of the 
isentropic expansion belong  to the Poisson adiabat at which 
$s/\rho_B = s^{FO} V^{FO}/ (2A) = const$. Here the  entropy  density $s^{FO}$ and the system 
volume $V^{FO}$ are taken at chemical FO, while the total number of baryons in
an  A+A collision is $2A$. At chemical FO temperatures $T$ below 150 MeV, the hadronic  
EOS can be safely represented as a mixture of ideal gases of massive  pions and nucleons, 
i.e., its pressure $p \simeq T (\rho_B + \rho_\pi) $  and energy density $\varepsilon \simeq 
(m_N + 3/2\,T) \rho_B + (m_\pi+3/2\,T)\rho_\pi $ can be represented via the 
density of nucleons $\rho_B$ and density of pions  $\rho_\pi$ (here $m_N$ ($m_\pi$) 
is the nucleon (pion) mass).  With the help of this EOS  the variable $X$ at chemical FO 
can be cast as $X^{FO} \simeq [m_N+m_\pi + 5/2 \, T(1+ \rho_\pi/\rho_B)] V^{FO}/ (2A) $.  
From Fig.\ \ref{fig5} one sees that  constant values of  $s/\rho_B$ 
in the range of $E_{lab} =  6.9-11.7$ GeV correspond to a nearly 
constant ratio $\rho_\pi/\rho_B \simeq 0.5$ and, hence, one can write  
$X^{FO} \simeq [m_N+m_\pi + 3.75 \, T] V^{FO}/ (2A) $ for these energies.  
Since for these energies the entropy density changes from 0.3 fm$^{-3}$ to 1.944 fm$^{-3}$, 
while the chemical FO temperature changes from 84 MeV to 127 MeV, it is clear that 
approximately one can   write  $X^{FO} \simeq [m_N+m_\pi + 358 \,{\rm MeV}] V^{FO}/ (2A) $ 
and, hence,  the value  $X^{FO} s^{FO} \simeq [1536 \, {\rm  MeV} ] V^{FO} s^{FO}/ (2A) = 
const$. 
A direct numerical check shows that for the chemical FO data belonging to the laboratory
energy range  $E_{lab} =  6.9-11.7$ GeV one obtains 
$X^{FO} s^{FO} \simeq 16.9; 19.3; 23.1$, which means that such a relation is 
valid with the relative deviations  $-12$\%  and  $+21$\%.
Note that the relation $X^{FO} s^{FO}/ [m_N+m_\pi + 3.75 \, T] \simeq const$ gives us the 
values $15; 16.5; 17.9$, i.e., it  is fulfilled  with relative  errors $-9.1$\% and   $+8.5$\% 
 in this energy range and these estimates validate our EOS usage.  
 Using these arguments, we conclude that with reasonable accuracy one can establish the 
 relation $X^{FO} \sim  V^{FO} \sim 1/ s^{FO}$ for the  final states which belong to the Poisson 
 adiabat and, therefore, the growth of entropy density (see Fig.\ \ref{fig1}) and the decrease of
 the variable $X$  shown in Fig.\ \ref{fig2} are directly  related to each other. 

The same treatment can be applied to higher energies. In this case one has to write $X^{FO} 
\simeq [m_N+m_\pi + 5/2 \, T(1+ \rho_\pi/\rho_B)] s^{FO}/\rho_B\, / s^{FO}$ and account for the 
fact that the ratios $s^{FO}/\rho_B$ and $\rho_\pi/\rho_B$ are increasing with the collision  
energy, while the chemical FO temperature and entropy density are almost constant for  
$E_{lab} > 11.7$ GeV (see Fig.\ \ref{fig1}).  Hence, in this case one can write $X^{FO} \simeq 
s^{FO}/\rho_B$, i.e.,  for laboratory energies above 11.7 GeV the variable  $X^{FO}$ should 
increase with the collision energy, and this is a reflection of the growth of the initial values of the 
$X$ variable, when the generalized shock adiabat goes inside the QGP. 

The same conclusion can be obtained from the fact that the Poisson adiabats with the different 
$s/\rho_B$ values cannot intersect each other. 
Therefore, the generalized shock adiabat  which must replace the unstable RHT adiabat  
(like the one shown in Fig.\ \ref{fig3}) would    generate the field of  nonintersecting 
Poisson adiabats in the $X-p$ plane, since along the mechanically stable hydrodynamic 
solutions the entropy cannot decrease. By construction at given $X$  the Poisson adiabat  
with higher value of $s/\rho_B$ has higher pressure $p$. 
Then applying the chemical FO criterion $p =const$, which  within the error bars   is  
clearly seen in Fig.\ \ref{fig2}, to such a field of nonintersecting Poisson adiabats in 
the $X-p$ plane,  one observes that the higher values of  $s/\rho_B$ correspond to larger 
values of the variable $X^{FO}$ along the line $p =const$. 

Accounting for  the above  estimates, we conclude that the local minimum of the $X^{FO}$ variable is related to the minimum of the variable $X$ 
on the generalized shock adiabat existing at the boundary between the mixed phase and QGP. 
Moreover, the above estimates show that the minimum of the $X^{FO} (E_{lab})$ function  
corresponds to the minimum of the chemical freeze-out volume $V^{FO} (E_{lab})$, reported in  
\cite{KABAndronic:05} and reanalyzed recently in  \cite{Oliinychenko:12}. Thus, we find that  
the minimum of  $V^{FO} (E_{lab})$ is 
generated by the unstable part of the RHT adiabat to the boundary of mixed and QGP phases, 
i.e., it is another signal of QGP formation.

Note that these conclusions  were  also verified numerically for the shock adiabat shown in 
Figs.\ \ref{fig3} and   \ref{fig5} and, hence, it is appropriate to present here the employed EOS.
The  hadron gas  pressure  used in the present work  accounts for 
the mesonic and the (anti)baryonic states which are described by the masses $m_M$, $m_B$ 
and by the temperature-dependent numbers of degrees of freedom
\cite{KABPoS12}
\begin{eqnarray}
\label{EqIV}
p_H &=& \left[  C_B T^{A_B}e^{-\frac{m_B}{T}} \cdot2\cosh\left({\mu}/{T}\right) 
+   
        C_M T^{A_M}e^{-\frac{m_M}{T}}\right]\cdot e^{-\frac{p_HV_H}{T}}\,.        
\end{eqnarray}
This 
EOS  accounts for the short-range repulsion introduced via the excluded  volume $V_H=
\frac{4}{3}\pi R_H^3$  (with $R_H = 0.3$ fm) taken to be equal for  all hadrons. With the 
parameters $m_M=8$ MeV, $m_B=800.5$ MeV and 
\begin{eqnarray}
A_M& =&4.95 \quad\quad C_M=6.90\cdot10^{-9}~{\rm MeV}^{1-A_M}{\rm fm}^{-3}\,, \nonumber \\
A_B & = & 6.087, \quad C_B=2.564\cdot10^{-9}~{\rm MeV}^{1-A_B}{\rm fm}^{-3}\,,\quad 
%%%m_M&=&7 ~{\rm MeV},  ~ m_B= 800.5 ~{\rm MeV}\,.   \nonumber 
%
\end{eqnarray}
such a model not only  represents the mass-integrated spectrum of all hadrons,  but  also  it  
rather  accurately reproduces the chemical FO densities of mesons $\rho_M$ and baryons $
\rho_B$ and the ratios $s/\rho_B$ and $s/\rho_M$ for the chemical FO temperatures below 155 
MeV \cite{KABPoS12}. The parameters of the center of the shock adiabat were fixed  as: 
$p_0=0$, $\rho_0=0.159\,\, {\rm fm}^{-3}$
and $\varepsilon_0=126.5~{\rm MeV~fm}^{-3}$.

The  QGP EOS is motivated by the MIT-Bag model \cite{MITBag}
$
p_Q=A_0T^4+A_2T^2\mu^2+A_4\mu^4-B 
$, 
where the constants  $A_0 \simeq 2.53 \cdot 10^{-5}~{\rm MeV}^{-3}{\rm fm}^{-3}$, $A_2 
\simeq 1.51 \cdot 10^{-6} ~{\rm MeV}^{-3}{\rm fm}^{-3}$, $A_4 \simeq 1.001 \cdot 10^{-9}~{\rm 
MeV}^{-3}{\rm fm}^{-3}$, and $B \simeq 9488~{\rm MeV}~{\rm fm}^{-3}$ were found by  fitting  
the  $s/\rho_B$  chemical FO data   for $E_{lab} <  50$ GeV with $s/\rho_B$ values  along the 
RHT adiabat   and by keeping the  pseudocritical temperature value 
 at zero baryonic density  close  to 150 MeV which is known from lattice QCD \cite{Aoki}. 
The phase diagram was found from the Gibbs criterion, $p_H(T,\mu_B) = p_Q(T,\mu_B)$.
The resulting RHT adiabat describes the  $s/\rho_B$  chemical FO data  well (see Fig.\ \ref{fig5}). 

\section{Discussion  of the results}
We have presented remarkable irregularities at chemical FO elucidated via a high-quality fit of 
experimental  particle ratios obtained  by the advanced version of the hadron resonance gas 
model. The achieved value of fit quality    $\chi^2/dof \simeq 1.16$ gives us a high confidence 
in our findings. Among these irregularities we observed a dramatic jump of the effective number 
of degrees of freedom and a local minimum of the generalized specific volume $X^{FO} 
(\sqrt{s_{NN}})$ at  center-of-mass collision energies  $\sqrt{s_{NN}} = 4.3-4.9$ GeV.  Also, at 
chemical FO we found plateaus in the collision-energy dependence of the entropy per baryon, 
of the  total and of the thermal  numbers of pions per baryon, which were predicted long ago 
\cite{KAB:89a, KAB:90, KAB:91}.  We discussed the generalized shock adiabat model for low 
energy collisions and argued that the found plateaus and the minimum of  $X^{FO} 
(\sqrt{s_{NN}})$ are generated by the RHT adiabat instabilities existing at the boundary 
between the mixed phase and QGP.  The numerical simulations of the RHT adiabat   for the 
realistic EOS of the hadronic phase allowed us to reproduce the
 $s/\rho_B$ plateau and to fix the parameters of the QGP EOS. Also, at chemical FO  we found 
 a second set of plateaus at $E_{lab} \simeq 30-40$ GeV, which, however,  do not correspond 
 to a phase transition or to the discussed instabilities. To make more definite conclusions about 
 the found plateaus at laboratory energy $30-40$ GeV we need more precise data measured 
 with  $E_{lab}$ steps of about 100-200 MeV.

\acknowledgments 
The authors are thankful to I. P. Yakimenko for valuable comments. 
This publication is based on the research provided by the grant support of the State Fund for Fundamental Research (project N  {\cyrfnt F}58/175-2014).
K.A.B.  and I.N.M.  acknowledge  a partial support provided by the Helmholtz 
International Center for FAIR within the framework of the LOEWE 
program launched by the State of Hesse. 
D.R.O. acknowledges funding of a Helmholtz Young Investigator Group VH-NG-822
from the Helmholtz Association and GSI, and   thanks HGS-HIRe for a support.
A partial support by the grant NSH-932.2014.2 
is  acknowledged by  I.N.M. and L.M.S.

%\clearpage

\end{document}